\documentclass[conference]{IEEEtran}
\IEEEoverridecommandlockouts
\usepackage{cite}
\usepackage{amsmath,amssymb,amsfonts}
\usepackage{algorithmic}
\usepackage{graphicx}
\usepackage{subcaption}
\usepackage{textcomp}
\usepackage{xcolor}
\usepackage{hyperref}
\usepackage{multicol}
\usepackage{dblfloatfix}
\usepackage{float} 
\def\BibTeX{{\rm B\kern-.05em{\sc i\kern-.025em b}\kern-.08em
    T\kern-.1667em\lower.7ex\hbox{E}\kern-.125emX}}
\begin{document}

\title{On the Design of Ethereum's Data Availability Sampling: A Comprehensive Simulation Study
}

\author{\IEEEauthorblockN{1\textsuperscript{st} Arunima Chaudhuri}
\IEEEauthorblockA{\textit{Codex} \\
India \\
arunima@status.im}
\and
\IEEEauthorblockN{2\textsuperscript{nd} Sudipta Basak }
\IEEEauthorblockA{\textit{Codex} \\
India \\
sudipta@status.im}
\and
\IEEEauthorblockN{3\textsuperscript{rd} Csaba Kiraly}
\IEEEauthorblockA{\textit{Codex} \\
Italy \\
csaba@status.im}
\and
\IEEEauthorblockN{4\textsuperscript{th} Dmitriy Ryajov}
\IEEEauthorblockA{\textit{Codex} \\
Costa Rica \\
dryajov@status.im}
\and
\IEEEauthorblockN{5\textsuperscript{th}  Leonardo Bautista-Gomez}
\IEEEauthorblockA{\textit{Codex, MigaLabs} \\
Spain \\
leo@status.im}
}

\maketitle

\begin{abstract}
This paper presents an in-depth exploration of Data Availability Sampling (DAS) and sharding mechanisms within decentralized systems through simulation-based analysis. DAS, a pivotal concept in blockchain technology and decentralized networks, is thoroughly examined to unravel its intricacies and assess its impact on system performance. Through the development of a simulator tailored explicitly for DAS, we embark on a comprehensive investigation into the parameters that influence system behavior and efficiency.

A series of experiments are conducted within the simulated environment to validate theoretical formulations and dissect the interplay of DAS parameters. This includes an exploration of approaches such as custody by row, variations in validators per node, and malicious nodes.

The outcomes of these experiments furnish insights into the efficacy of DAS protocols and pave the way for the formulation of optimization strategies geared towards enhancing decentralized network performance. Moreover, the findings serve as guidelines for future research endeavors, offering a nuanced understanding of the complexities inherent in decentralized systems.

This study not only contributes to the theoretical understanding of DAS but also offers practical implications for the design, implementation, and optimization of decentralized systems.

\end{abstract}


\section{Introduction}

In recent years, the rise of decentralized systems, especially in blockchain, has highlighted challenges in scaling transaction throughput while maintaining data availability and integrity across the network. Among the techniques developed to address these challenges, Data Availability Sampling (DAS) \cite{b6, b8} has emerged as a promising approach. DAS is crucial in data sharding \cite{b1, b14, b15}, where large datasets are divided into smaller fragments termed "data-shards" \cite{b25, b26}. Each node stores only a portion of the dataset, enabling parallelization and distribution of data storage and computation.

Danksharding \cite{b2, b3, b6} is a sharding approach proposed for Ethereum 2.0 that extensively uses DAS. It aims to enhance scalability \cite{b10} by introducing "blobs" to store transactional data alongside core blockchain data. By efficiently managing data availability for these blobs using DAS, danksharding seeks to create a highly scalable \cite{b21, b22} and secure Ethereum \cite{b13} network.

DAS aims to verify the availability and correctness in a decentralized network, especially in data sharding environments by providing a sampling based mechanism for nodes to verify, with a high probability, the availability of the dataset, even with partial data views. For this to work, the data has to be encoded first using erasure coding, before being distributed in the network. 

First, data is split into fragments by the block producer. Each block of data is represented as a matrix, where the original data fills the cells. In the context of data sharding, a 2D structure refers to arranging the data like a table with rows and columns. Erasure coding is then applied horizontally (across each row) and vertically (across each column) to create additional redundant data points. These redundant data points, along with the original data, allow for the recreation of missing pieces if they become unavailable. This approach offers high data reliability because even if cells (data samples) are missing, these can be recovered using the corresponding row or column parity data. 

Instead of receiving the entire block, nodes in the network are given only a few rows and columns of data. These are distributed across the network using techniques like gossipsub \cite{b4, b5, b7, b12} channels, which implement publisher/subscriber \cite{b11, b18, b19} dissemination, reducing message overhead compared to traditional flooding methods \cite{b9}. Validators attest to the availability and correctness of a block based on the rows and columns they receive.

If a node fails or data is corrupted, missing data can be reconstructed using the remaining fragments and parity blocks. This enables efficient data repair without replicating the entire dataset across the network. DAS ensures data availability, while erasure coding facilitates data repair, maintaining the integrity and resilience of the stored data. DAS enables nodes to assess data availability without needing the entire dataset, allowing nodes with limited resources (bandwidth and storage capacity) to participate, thus democratizing involvement in decentralized networks.

The interplay between data sharding and DAS introduces complexities that require thorough exploration for designing decentralized systems. As the Ethereum community and other blockchain projects scale\cite{b20} and adopt data sharding, robust simulation tools are needed to understand DAS protocols. To address these challenges, this paper offers a detailed study of DAS and sharding through simulation-based analysis. We introduce a simulator \cite{b23} designed to analyze the impact of various parameters on system performance.

The remainder of this paper is organized as follows: Section \ref{section_background} provides background on DAS and its role in decentralized systems, emphasizing the need for simulation-based analysis. Section \ref{section_simulator} details our simulator, including its parameters, implementation, and significance. Section \ref{section_evaluation}presents the results of our experiment, offering a detailed evaluation of the outcomes.  Section \ref{section_analysis} discusses the conclusions drawn from the experiment. Finally, Section \ref{section_conclusion} summarizes our findings and suggests directions for future research.

\section{Background}
\label{section_background}

DAS is employed in blockchain, particularly with data sharding, to verify the availability of data across a decentralized network. Blockchains can be sharded in various ways, either by dividing both data and computation or by sharding data alone while keeping execution distributed. Ethereum's data sharding proposal \cite{b21} focuses on sharding data availability, ensuring that not every node stores the entire blockchain.

Data sharding is crucial for accommodating nodes with limited storage while maintaining network security. The challenge is ensuring data integrity when nodes have only partial views of the dataset. DAS addresses this in the context of Ethereum by enabling nodes to verify the availability of the full dataset through:

\begin{itemize}
\item Dissemination: Initially, a certain number of validators in the network are provided with rows and columns of data, attesting to the availability and correctness of the block.
\item Sampling: Nodes in the network randomly query other nodes to retrieve different samples of data. These samples are then used to check whether the entire data block is available or not.
\item Verification: By querying a sufficient number of random samples and receiving them correctly, nodes can gain confidence in the availability of the entire data block. DAS allows for a probabilistic verification process, where a high number of correct samples implies a high probability of the block being available.
\end{itemize}

The interplay between data sharding and DAS is complex and requires in-depth exploration and analysis. The Ethereum Foundation acknowledges the necessity for a simulator to untangle these complexities. The simulator serves as a virtual playground, where the implications of various configurations can be observed and understood. This is crucial in shaping the design and deployment of decentralized systems. This simulator is a valuable tool for the Ethereum community and beyond, offering insights into decentralized network behavior in various conditions.

\section{Simulator}
\label{section_simulator}

The open-source simulator\footnote{Simulator available at \url{https://github.com/codex-storage/das-research}} is written in Python to facilitate modification by the community. It supports the parallel execution of large-scale simulation studies, with parameter space exploration and multiple repetitions for statistical relevance. It also supports reproducible simulations and the automated plotting of simulation results and summary statistics.

The simulator's configuration is specified through a Python script, offering a flexible approach to define simulation settings. Each parameter offers researchers control over the simulation environment, allowing them to explore and analyze the dynamics of decentralized systems under diverse conditions. The following sections provide a brief overview of selected parameters. For more details we refer the reader to the source repository.

\subsection{Block and DAS Configuration:}
\begin{itemize}
\item Row/Column Size N: Number of columns/rows after erasure coding, configurable separately for each dimension.
\item Row/Column Size K: Erasure coding parameters, with K unique cells in a row(columns) out of N required to recover all N cell in the row(column). Erasure coding can be set independently for each dimension, allowing also 1D models where only one of the dimensions have $N>K$.
\item custodyRow/Column: Number of rows/columns a node or validator is interested in. Can be set to 0 for one dimension to simulate 1D configuration.
\item cellSize: the size of a cell in the erasure coding matrix, in bytes.
\end{itemize}

\subsection{Network and Overlay Configuration:}
\begin{itemize}
\item Number of nodes: total number of full nodes with validators in the overlay network.
\item class1ratios: Ratio of class 1 nodes. Class 1 nodes represent small solo stakers and class 2 nodes represent large staking operator nodes.
\item vpn1 and vpn2: Number of validators per node (vpn) for class 1 and class 2 nodes, respectively.
\item Network Degree: Per-topic gossipsub mesh neighborhood size.
\end{itemize}

\subsection{Failure Model Configuration:}
\begin{itemize}
\item Failure Rate: Percentage of data not released by the block producer.
\item Malicious Nodes Rate: Percentage of nodes considered malicious and do not forward any data.
\end{itemize}

\subsection{Network Underlay Configuration:}
\begin{itemize}
\item Producer, Class1, Class2 (upload b/w): Uplink bandwidth in megabits/second of the block producer, class 1 nodes and class 2 nodes, respectively.
\item Latency: Network latency on peer-to-peer links. Note that additional bandwidth dependent queuing latency is considered on top of this underlay latency.
\end{itemize}

\section{Evaluation}
\label{section_evaluation}

This section details a series of experiments (in subsection \ref{subsection_exp1}) to investigate the impact of the number of rows and columns assigned to each validator on the total delivered samples. We also explore how the FullDAS \cite{b17, b24} principles for enhancing network robustness and scalability are supported by these experiments.

The theoretical calculation for the total number of samples to deliver is:

\begin{equation}
\begin{split}
\text{Total samples to deliver} &=  (N-1) \times  \text{class1ratio} \times \text{vpn1} \\
&\quad \times (\text{RowSize} \times \text{CustodyRow} \\
&\quad + \text{ColumnSize} \times \text{CustodyColumn}) \\
&\quad + (1 - \text{class1ratio}) \times \text{vpn2}  \\
&\quad \times (\text{RowSize} \times \text{CustodyRow} \\
&\quad + \text{ColumnSize} \times \text{CustodyColumn})
\end{split}
\label{equation}
\end{equation}

\begin{itemize}
\item N: Total number of nodes.

\item class1ratio: Ratio of Class 1 nodes to total number of nodes.

\item RowSize and ColumnSize: In the context of data sharding and DAS, rows and columns refer to the units used for partitioning data within the network. RowSize and ColumnSize parameters denote the sizes of rows and columns, respectively.

\end{itemize}

The distinction between custody and vpn lies in their respective responsibilities:

\begin{itemize}

\item CustodyRow and CustodyColumn: Custody, in the context of our simulation, refers to the responsibility of nodes for storing and validating specific data partitions (rows and columns) within the decentralized network. CustodyRow and CustodyColumn represent the number of rows and columns nodes have the custody of. Nodes entrusted with custody are responsible for ensuring the availability and integrity of the data assigned to them, contributing to the overall reliability of the network.

\item vpn1 and vpn2: Validators per node (vpn) represent the number of validators assigned to each individual node. The parameter vpn1 denotes the number of validators per node for Class 1 nodes, while vpn2 represents the corresponding value for Class 2 nodes. Nodes may have multiple validators who are assigned rows and columns to custody. Under this construct, the node has to custody the union of the rows and columns of all the validators it hosts.

\end{itemize}

While validators per node focus on transaction validation and consensus-building, custody pertains specifically to data storage and validation within the context of data sharding mechanisms. 

\begin{table}[h]
\centering
\caption{Simulator Configurations}
\label{tab:simulator_configurations}
\begin{tabular}{|l|l|}
\hline
\textbf{Parameter}                 & \textbf{Value}          \\ \hline
Number of Nodes                    & 5000                    \\ \hline
Row, Column Size                   & 100, 100                \\ \hline
Row, Column Size (K)               & 100, 100                \\ \hline
vpn1, vpn2                         & 1                        \\ \hline
Class1 Ratio                       & 0.8                     \\ \hline
Producer, Class1, Class2 (upload b/w) & 200, 10, 200            \\ \hline
Network Degree                     & 8                       \\ \hline
Failure Rate, Malicious Rate       & 0\%                     \\ \hline
\end{tabular}
\end{table}

\subsection{Experiments to explore the Impact of Rows and Columns per Validator on Total Delivered Samples}
\label{subsection_exp1}

\begin{figure}[htbp]
\begin{subfigure}{0.5\textwidth}
\includegraphics[width=\textwidth]{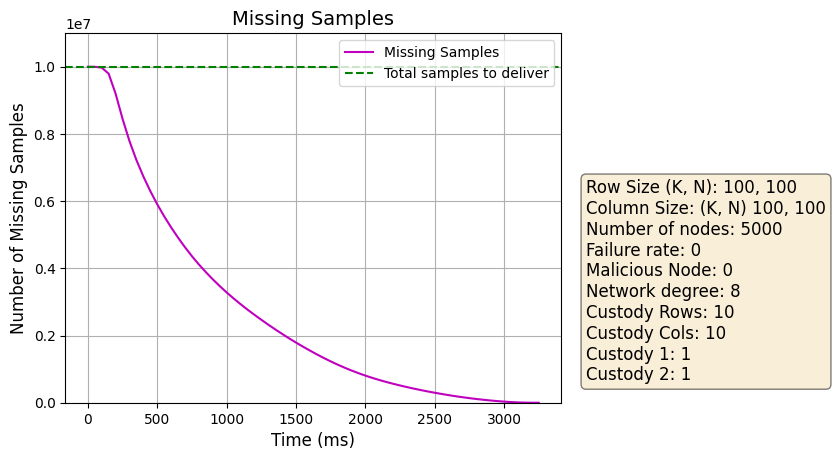}

\label{fig:sub10}
\end{subfigure}
\caption{Progress of the number of samples still missing from custody, as a function of slot time (rows/columns per validator = 10)}
\label{fig:exp_1}
\end{figure}

To empirically validate the formula \ref{equation}, we conducted experiments by varying the custody of rows and columns per validator (1, 2, 3, 4, 5, 10, 50, 100). The configurations used for this experiment are specified in Table \ref{tab:simulator_configurations}. The green line in the plots \ref{fig:exp_1} signifies the theoretical value, serving as a reference for understanding deviations from the expected outcome, while the purple line illustrates the changes in missing samples throughout the length of the simulation. It shows the time it takes to propagate all the samples in a network with 5000 nodes.

\begin{table}[h]
\centering
\caption{Results}
\label{tab:experiment1_results}
\begin{tabular}{|c|c|c|c|c|}
\hline
\textbf{Custody} & \textbf{Custody} & \textbf{Observed} & \textbf{Theoretical} & \textbf{Difference} \\ 
\textbf{Row} & \textbf{Column} & & & \\ \hline
1 & 1 & 999800 & 999800 & 0 \\ \hline
2 & 2 & 1999600 & 1999600 & 0 \\ \hline
3 & 3 & 2999400 & 2999400 & 0 \\ \hline
4 & 4 & 3999200 & 3999200 & 0 \\ \hline
5 & 5 & 4999000 & 4999000 & 0 \\ \hline
10 & 10 & 9998000 & 9998000 & 0 \\ \hline
50 & 50 & 49990000 & 49990000 & 0 \\ \hline
100 & 100 & 99980000 & 99980000 & 0 \\ \hline
\end{tabular}
\end{table}

Based on Table \ref{tab:experiment1_results} and Figure \ref{fig:exp_1}, the experimental results demonstrate close agreement between observed and theoretical values across various configurations of custody rows and columns.

This consistency validates the proposed formula \ref{equation} for estimating the total number of samples to deliver. This reveals insights into how different custody allocations impact sample delivery, supporting FullDAS's recommendation for strategic custody allocation to optimize network performance.

\section{Analysis}
\label{section_analysis}

This section presents analysis of the experiment and examines the key takeaways and their implications for Ethereum DAS design and implementation in other decentralized networks.

\subsection{Impact of custody rows and custody columns on total sample delivery}

The experiment conducted using the simulator examined the influence of various configurations of custody rows and custody columns on the total delivered samples. The close agreement between observed and theoretical values (Table \ref{tab:experiment1_results}, Figure \ref{fig:exp_1}) across various configurations validates the proposed formula \ref{equation} for estimating sample delivery. Thus system designers can predict the performance of DAS systems accurately based on different custody configurations and optimize network parameters to achieve desired performance levels by tweaking custody rows and columns.

Overall, this research guides the Ethereum Foundation in the development of scalability solutions for Ethereum, making it a significant contribution to the blockchain community.

\section{Conclusion}
\label{section_conclusion}

In conclusion, this paper provides a comprehensive examination of DAS. Our efforts start with the development of a simulator for conducting experiments related to DAS protocols. We validate theoretical formulations, analyze the impact of various parameters on system performance, and evaluate the effectiveness of the simulator itself.

Our findings contribute to the understanding of DAS and its implications for Ethereum scalability. By systematically varying parameters such as block size, network degree, validators per node, custody configurations, and malicious nodes we gained valuable insights into the factors influencing the efficiency and reliability of DAS protocols.

The simulator presented in the paper is being used to gain insight into expected system behaviour and performance characterization for the upcoming PeerDAS~\cite{b16} improvement foreseen to be introduced in the Pectra fork of Ethereum. Improved versions of the simulator will also serve as baseline for evaluating further improvements and novel techniques outlined in FullDAS~\cite{b24}.

Our work also lays the foundation for further research in distributed data availability protocols and their applications in decentralized systems. The insights gained from our experiments provide a roadmap for future investigations, including the exploration of more complex network topologies, the integration of additional security mechanisms, and the scalability of DAS protocols to larger networks and datasets. By combining theoretical analysis with practical experimentation, we offer valuable contributions to the ongoing research on DAS, which is critical for the scalability of Ethereum and other blockchains.

\section{Acknowledgements}
\label{section_acknowledgements}

This work was supported by the Ethereum Foundation under grant FY22-0820. We also would like to thank Dankrad Feist and Danny Ryan for many discussions and important feedback.

\end{document}